\documentclass{article}
\usepackage[utf8]{inputenc}
\usepackage{graphicx}
\usepackage{authblk}
\usepackage{hyperref}
\usepackage[a4paper]{geometry}
\begin{document}

\title{Low-energy Ion Beam Diagnostics: An Integrated Solution}
\author[1,10]{A.Ad{\i}g\"{u}zel}
\affil[1]{\.{I}stanbul University, Department of Physics, \.{I}stanbul}
\author[2]{H.\c{C}etinkaya}
\affil[2]{K\"{u}tahya Dumlup{\i}nar University, Department of Physics, K\"{u}tahya}
\author[1]{\c{S}.Esen}
\author[3]{D.Halis}
\affil[3]{Y{\i}ld{\i}z Technical University, Department of Physics,  \.{I}stanbul}
\author[4]{T.B.\.{I}lhan}
\affil[4]{\.{I}stinye University, Department of Physics,  \.{I}stanbul}
\author[5]{A.K{\i}l{\i}\c{c}gedik}
\affil[5]{Marmara University, Department of Physics,\.{I}stanbul}
\author[6]{S.O\u{g}ur}
\affil[6]{ADAM SA, Geneva}
\author[7]{S.\"{O}z}
\affil[7]{Bo\u{g}azi\c{c}i University, Department of Mechanical Engineering, \.{I}stanbul}
\author[8]{A.\"{O}zbey}
\affil[8]{\.{I}stanbul University Cerrahpa\c{s}a, Department of Mechanical Engineering, \.{I}stanbul}
\author[9,10]{V.E.\"{O}zcan}
\affil[9]{Bo\u{g}azi\c{c}i University, Department of Physics, \.{I}stanbul}
\author[11,9]{N.G.\"{U}nel}
\affil[11]{University of California Irvine, Physics Department, Irvine}
\affil[10]{Bo\u{g}azi\c{c}i University, Feza G\"{u}rsey Center for Physics and Mathematics, \.{I}stanbul}
\date{ \today }
\maketitle

\begin{abstract}
High gradient accelerator injectors have been widely studied throughout the world-leading accelerator facilities. The demand for high frequency cavities have led the Detector, Accelerator and instrumentation laboratory (KAHVELab) in Istanbul to deploy a four-vane Radio Frequency Quadrupole (RFQ) operating at 800~MHz to accelerate 20~keV protons to 2~MeV. The protons from the microwave discharge ion source  are transversely matched to the RFQ via an optimized Low Energy Beam Transport (LEBT) line which also contains an integrated measurement station, called measurement station (MBOX). The MBOX is used to measure the proton beam’s current along with pulse length duration, profile as well as the beam emittance upstream of the RFQ. It contains a number of home-built diagnostic tools: a Faraday cup, a scintillator screen and a pepper pot plate (PP). The analysis software is also locally developed and tested for the PP photo analysis.
In this note, the design, construction and tests of the integrated measurement station are discussed. The results from various measurements, especially on beam profile and charge, are compared to the simulation predictions.

\end{abstract}

\section{Introduction}

In an ion beamline, it is essential to be able to measure the properties of the beam, especially before it enters an accelerating cavity. This is even more critical for a low energy ion beamline as the Radio Frequency Quadrupole, which could be damaged by an unmatched beam, is costly in terms of financial resources and time. 
Therefore, measuring the ion beam properties (such as its current, profile and emittance) is critical for its acceleration and its downstream use. This makes an accurate and effective diagnostic station a necessity for the entire beamline. An additional example can be found in the field of nuclear medicine: it is very important to correctly measure the cross-sectional area of the beam, the number of particles passing through (beam current) and the beam emittance to determine the area and depth of the penetrating radiation that will define the treatment.

Here, we present our measurement station, custom designed and built based on previous experiences \cite{SPPpaper}.
The measurement station is currently installed on the proton test beam at KAHVELab, Istanbul, Turkey \cite{KAHVELab}.
The proton beamline is being constructed as part of an ongoing R\&D program with the short term goal of obtaining a PIXE (Proton Induced Xray Emission) measurement setup. Regarding the PIXE's efficiency onto lower Z-materials, i.e. Z=11 to 32, and lower impact onto the studied material, building a high frequency RFQ enabling 2 MeV beam energy with a charge at the order of $\mu$A is of great usage \cite{Tessa}. Furthermore, the long-term goals of the project range from building a medical proton accelerator to production of ion implant machines for semiconductor production. A generic goal of this R\&D program is to use domestic resources as much as possible, for the design and production of related parts (such as the beamline components, magnets, etc.) to increase the high technology awareness in the local manufacturers. Additionally, students and experts will be trained on the job, and considerable experience will be gained on both design and production on the beamline components and beam monitoring devices. Moreover, since the computer control of the whole system is to be carried out with home-grown software, the expertise gained in system control could also be reflected to other ongoing national projects such as the Turkish Accelerator Center \cite{TARLA}.

Therefore, the particular diagnostics box study presented in this note has two objectives: 1) to design and produce a test station for measuring the properties of the ion beam, and to develop the associated computer software to monitor and control the equipment. The control software, using some degree of automatization, is expected to minimize the errors and the needs for human intervention to the detector set while measuring the beam properties.
2) to use the developed hardware and software in a realistic beamline and to develop the necessary software for the analysis of the acquired data.
Accordingly, this paper starts by presenting the proton beamline and its components. The next section focuses on the details of the measurement station and the last section gives an outlook and conclusions.

\section{Proton Testbeam at KAHVELab}
\begin{figure}[!htbp]
\centering
\includegraphics[scale=0.4]{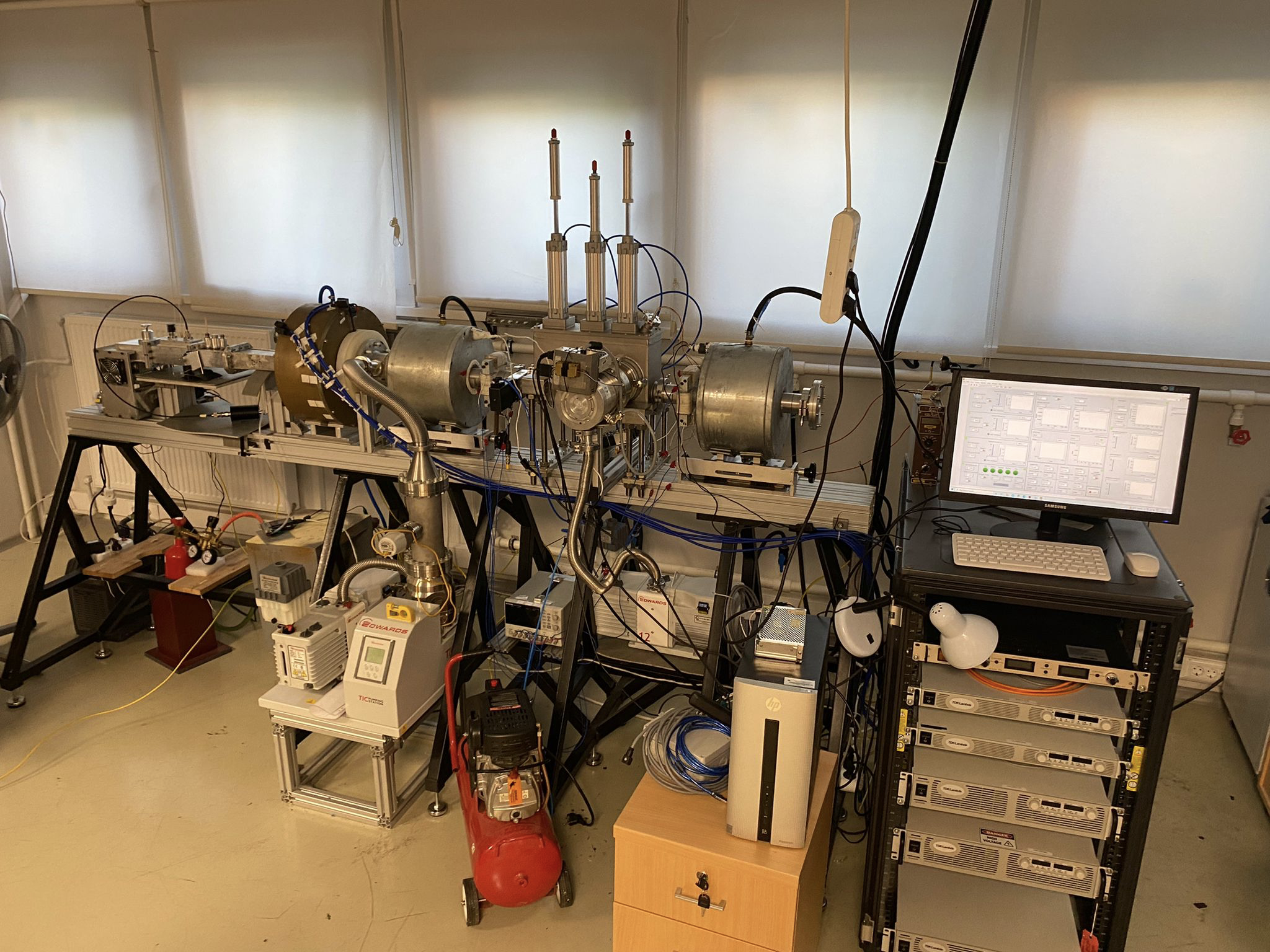}
\caption{The Proton Beamline at KAHVELab. RFQ is not included in this photo.}
\label{fig:beamline}
\end{figure}

The proton beamline design consists of an ion source, a low energy beam transport (LEBT) section and a radio frequency quadrupole (RFQ) at KAHVELab. Such a simple setup is the first stage of almost all modern ion accelerators such as the Linac4 injector of the Large Hadron Collider complex at CERN \cite{Linac4}. A view of the KAHVELab proton beamline installation can be seen in Fig. \ref{fig:beamline}.  

\subsection{Ion Source and LEBT Line}
A microwave discharge type ion source (MDIS) was selected for its low production cost, long lifetime,  high beam current, reliability,  low maintenance requirements. 
The MDIS setup consists of three main parts: The RF section, the plasma chamber and the extraction system, respectively. The details of MDIS design and production can be found elsewhere \cite{KAHVELabMDIS}.

\begin{figure}[!htbp]
\centering
\includegraphics[scale=0.5]{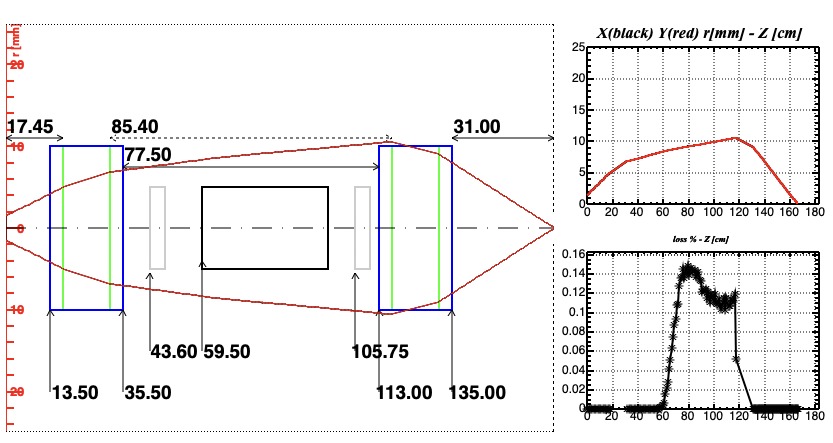}
\caption{LEBT Line designed by using TRAVEL and DemirciPro programs, image from DemirciPro. RMS beam envelope is shown in through the beam line and plotted also on right top figure with beam loss underneath, units are in cm.}
\label{fig:LEBTLine}
\end{figure}

After the Ion Source, the proton beam line continues with a Low Energy Beam Transport (LEBT) line. The aim of the LEBT line is to transmit the proton beam from the extraction point of the ion source to RFQ entrance point in a most efficient way.

The LEBT line shown in Fig. \ref{fig:LEBTLine}, was designed using TRAVEL\cite{travel} and DemirciPro\cite{demircipro} software programs. The solenoid effective length are imported into the dynamics code, where the solenoid effective length is calculated as: 
\begin{equation}
    \ell_{eff} =  \frac{\int_{z_1}^{z_2}{B_z^2 dz}}{B_{eff}^2}, 
\end{equation}
with $B_{eff}=B_{max} $  that refers to field at the solenoid center, i.e. the maximum longitudinal solenoid field component:~$B_z$. Regarding the hard edge solenoid assumption and the solenoid field map, this assumption has resulted a better agreement with the beam measurements than the direct interpretation of the Amp\`ere's law. In that figure, the beam moves from left to right. The leftmost vertical solid line represents the beam extraction from the IS and the rightmost dashed line represents the coupling to the RFQ. 
The total LEBT length is about 165~cm. 
Although the details of the LEBT line are presented elsewhere it is worth mentioning that the LEBT line has two movable solenoids for beam focusing (shown as blue rectangles) and two steerer magnets (shown as 
smaller gray rectangles) for beam guiding. A diagnostics station is placed between the two steerer magnets. 
The red line represents the RMS beam envelope minimized at the RFQ entrance.

\begin{figure}[!htbp]
\centering
\includegraphics[scale=0.4]{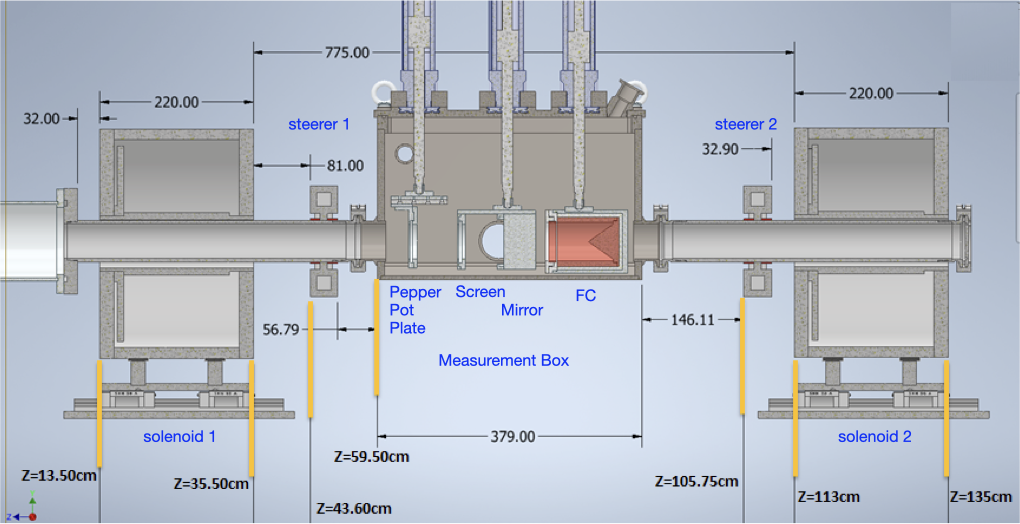}
\caption{LEBT Line designed via TRAVEL and DemirciPro programs, lengths in mm and positions in cm}
\label{fig:LEBTLineMbox}
\end{figure}

\begin{figure}[!htbp]
\centering
\includegraphics[width=0.80\linewidth,height=0.5\linewidth]{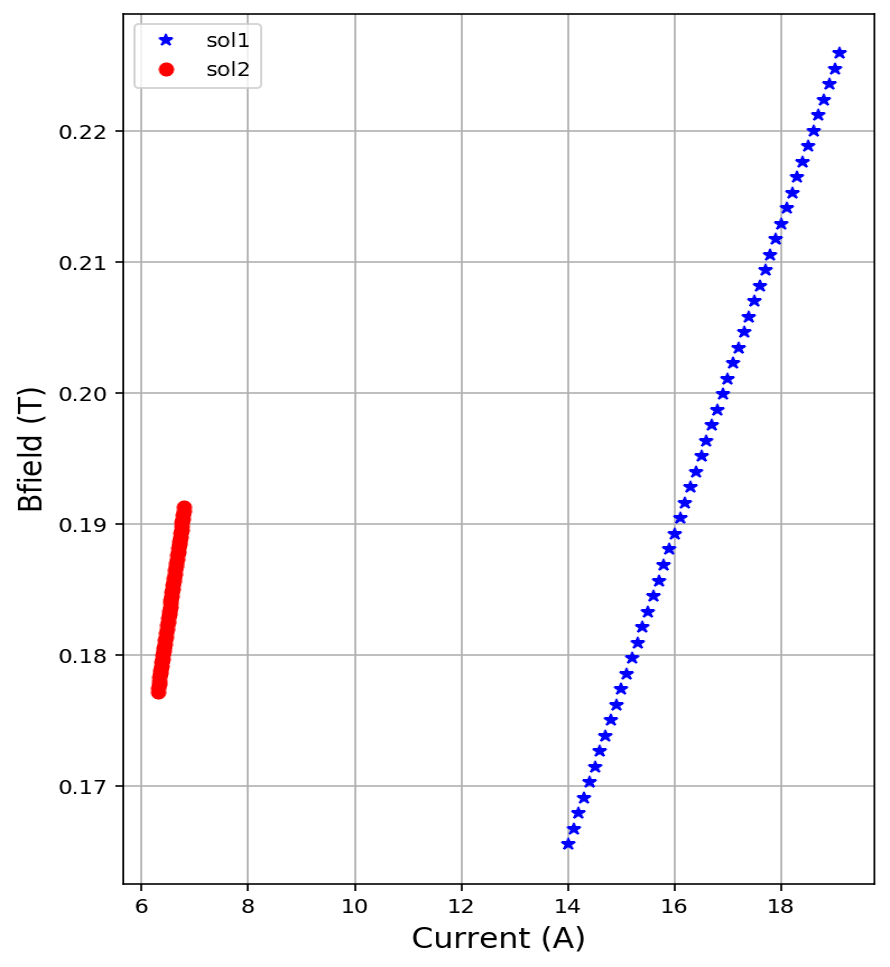}
\caption{Sol-1 (blue) and Sol-2 (red) magnetic fields versus the PSU current}
\label{fig:SolScan}
\end{figure}

The measurement results for a current scan on the focusing solenoids is shown in Figure \ref{fig:SolScan}. As expected, the first solenoid has a much wider magnetic field magnitude range to cover any imperfections of the ion source extraction system whereas the second solenoid allows a finer control over a much smaller range to match the RFQ input parameters. 

\subsection{RFQ}
The Radio Frequency Quadruople (RFQ) based on an in-house designed and to-be-locally-produced is the last part of the proton beamline. It operates at 800\,MHz to achieve the target energy. This RFQ is 4-vane cavity to accelerate the 20\,keV H+ beam extracted from plasma ion source to 2\,MeV. It has been optimized to accelerate a proton beam of 1\,mA, in a 98\,cm cavity with an acceleration efficiency of about 30\,\% with the lowest RF power. The details of the RFQ can be found elsewhere \cite{RFQpaper}.

\section{Diagnostic Station}

The characterization and the diagnostic of the ion beam is achieved by the so called ``measurement box'' (MBOX) installed between the two solenoid magnets in LEBT line as shown in Figure \ref{fig:LEBTLineMbox}.
The MBOX, contains three different detectors for beam charge, profile and emittance measurements which are the Faraday Cup (FC), Scintillation Screen (SS) and Pepperpot (PP) Plate. Furthermore, the FC is also used to acquire information on the pulse duration regarding the signal generated in the FC.

\begin{figure}[!htbp]
\centering
\includegraphics[scale=0.49]{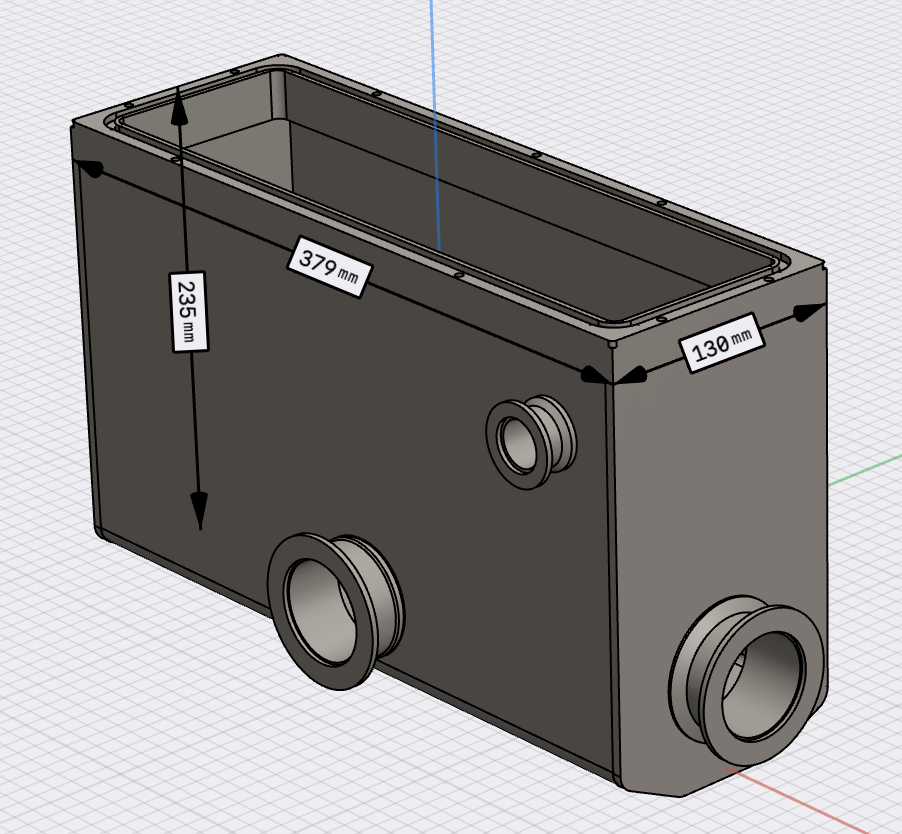}
$\quad$
\includegraphics[scale=0.20]{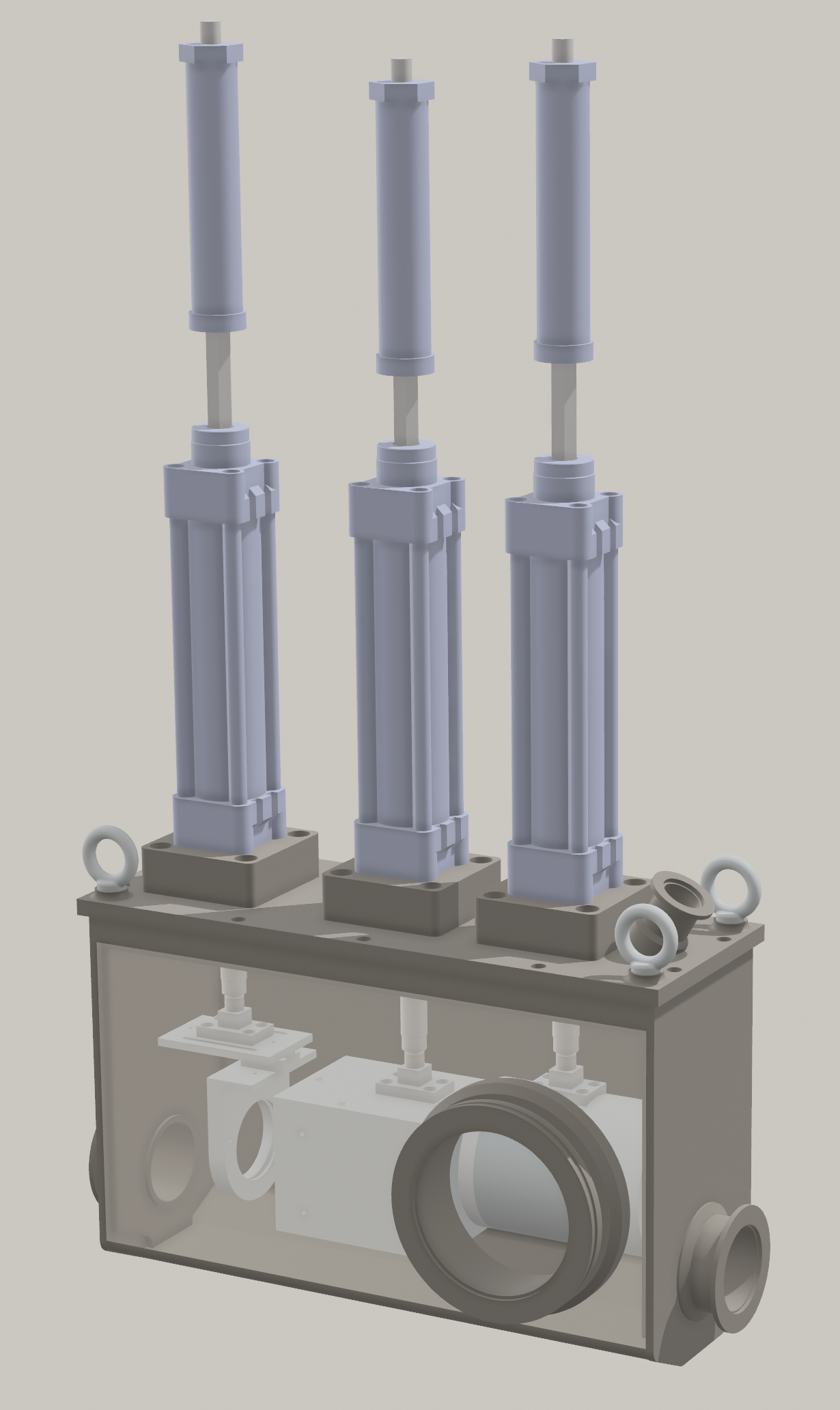}
\caption{Left: MBOX lower section design drawing, all units are in mm, Right: MBOX with actuators attached, the side wall is rendered transparent for clarity}
\label{fig:mboxProd}
\end{figure}

To keep the MBOX as compact as possible its 
vacuum vessel is designed and manufactured with the following two requirements: 1) the volume under vacuum  should be kept as small as possible but be large enough to host the relevant detectors 2) the wall thickness should be reduced as much as possible for a minimum total weight bit the vessel should be still robust enough to withstand the atmospheric pressure. 
The final outcome was a stainless steel  polygonal box of 5~mm wall thickness with dimensions of $379\times130\times235$~mm. The lower section of the box weighs about 15~kg and the top cover about 10~kg. 
The design drawings of the MBOX body can be seen in Figure \ref{fig:mboxProd} left side.
The connection to the beampipe is achieved with KF50 type connectors.
One lateral side of the MBOX houses the vacuum port, whereas the other one contains the glass view-port, and vacuum gauge port (KF40). Its upper cover houses a two channel feed-through connector (KF25) and three 3 pneumatic actuators to move the associated detectors into and out of the beamline. 
The Figure \ref{fig:mboxProd} right side shows the MBOX with the top cover and the actuators attached. In this image, the MBOX is shown from the other side and its lateral side is rendered transparent to show the detectors, the vacuum pump connector (ISO100) can also be seen. 
A remote-controlled gate valve is also installed upstream of the MBOX to separate the ionization chamber section from the beam diagnostics detectors, allowing for intervention with minimal disturbance to the beamline.
Inside the MBOX, the most upstream actuator is connected to the emittance meter, the middle one is attached to the screen and the plane mirror facing the view-port and the downstream to the Faraday Cup. 
The vacuum sealing between the lower and upper parts of the vessel is achieved with a Viton o-ring. With a single turbo-molecular vacuum pump, a pressure of $10^{-6}$~mbar was easily achieved using an independent setup as in Figure \ref{fig:mboxProd} right side.

\subsection{Detectors for Beam Diagnostics }
\begin{figure}[!htbp]
\centering
\includegraphics[scale=0.6]{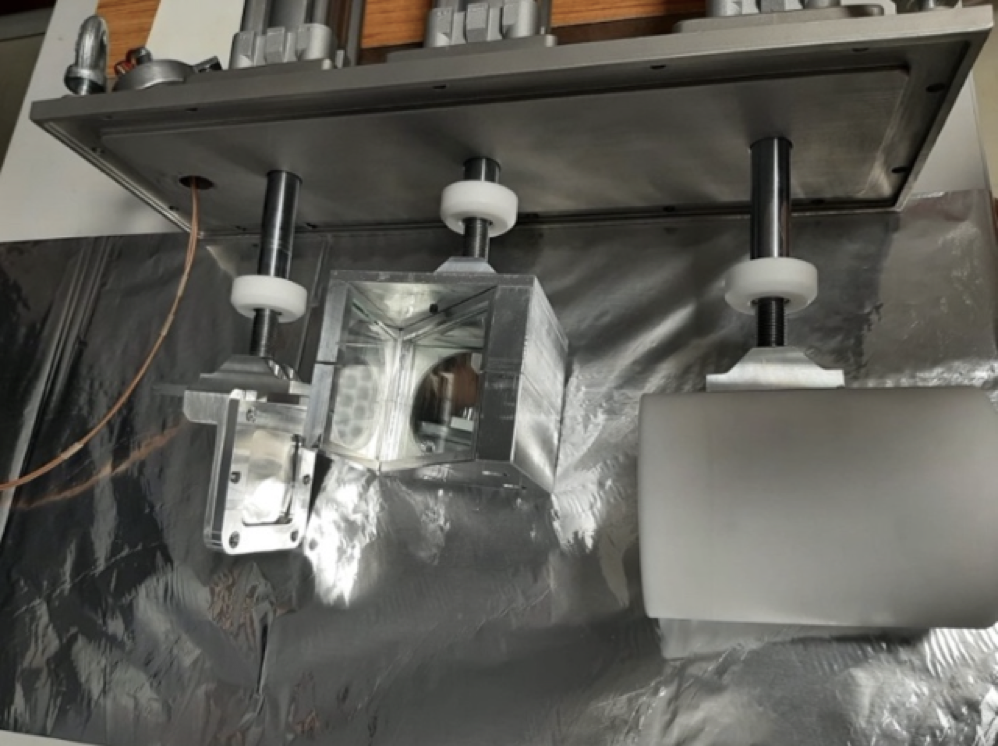}
\includegraphics[scale=0.14]{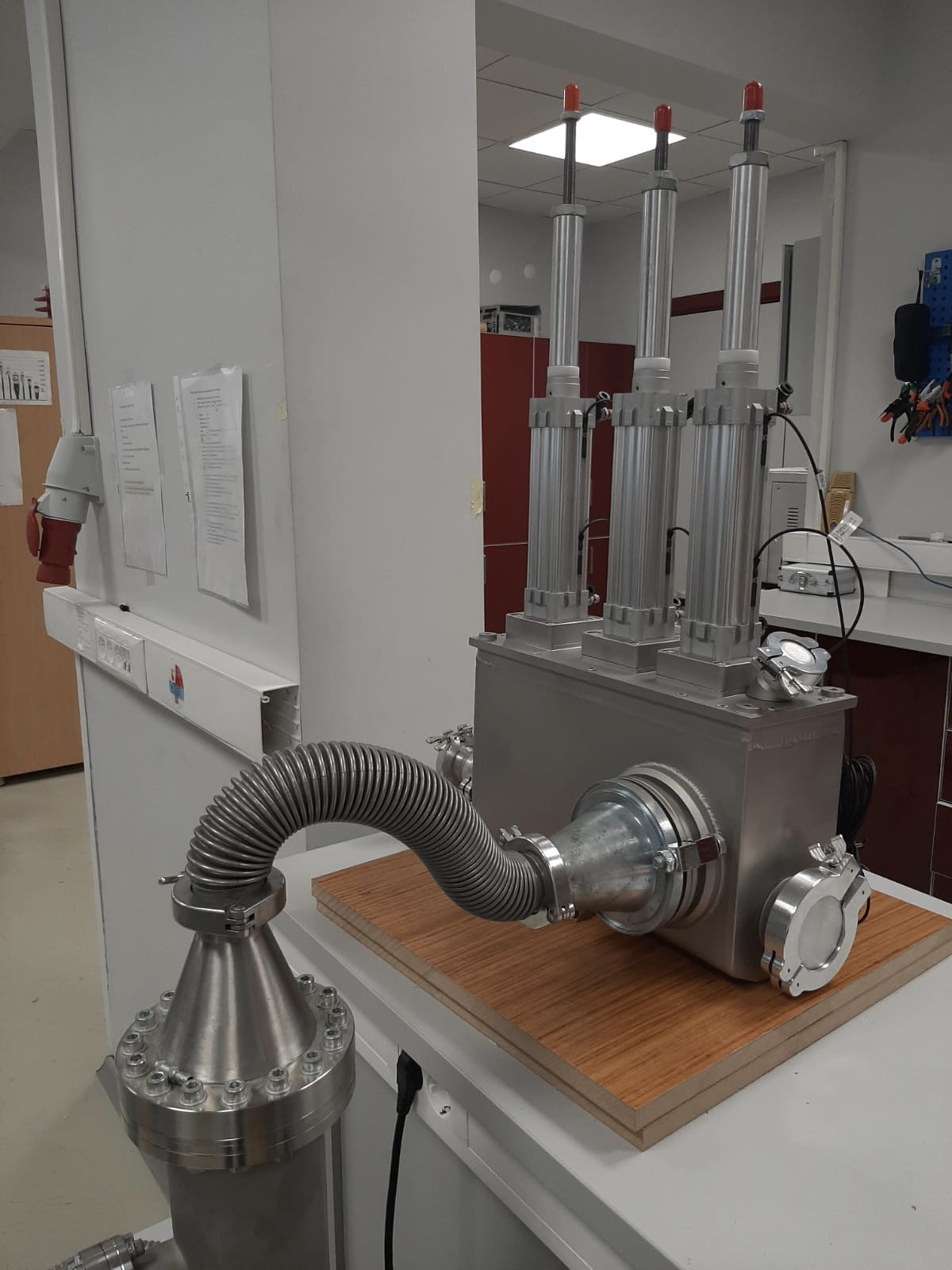}
\caption{Left: PP, SS and FC detectors for MBOX,  Right: MBOX vacuum test setup }
\label{fig:mboxDets}
\end{figure}

The term beam diagnostics usually refers to the measurement of the 6-Dimensional beam properties such that their dynamics can be constructed and beam parameters can be monitored through the accelerator. Particular to the KAHVELab LEBT line, diagnostics refers to the measurements of the beam transverse profile, beam emittance and the beam current as well as pulse duration. Emittance measurement being the peculiar one, it can be measured using various methods such as the quadrupole scan technique \cite{QStech}, slit scanning technique \cite{SStech}  or laser wire technique \cite{LWtech}.  
To benefit from past experience and to be able to measure both the X and Y components simultaneously, in our setup the pepperpot technique is used \cite{SPP_box}. 
Although this technique will be discussed in the following text, it should be mentioned that it has the advantage of monitoring the transverse beam emittance and profile in quasi-real time. 
The Figure \ref{fig:mboxDets} contains a photo of the three detectors of the MBOX as these are attached to the actuators discussed above. In this view, the beam moves from left to right and the detectors are the pepperpot plate (PP), the scintillator screen (SS) and the Faraday cup (FC). The remainder of this section discusses these detectors individually.

\subsubsection{Faraday Cup}
The beam current measurements are carried out using a Faraday Cup (FC). In the literature, Faraday cups can be classified into two groups: electrically biased or magnetically filtered. In this setup, an electrically biased Faraday cup was preferred to prevent reading errors that may originate from beamline magnetic elements. It is important to prevent the escape of secondary electrons and ions from Faraday cup to avoid either positive nor negative beam current readings \cite{FC1}.
The energy transferred from a 20 keV proton beam to secondary electrons as a result of Coulomb Collision is calculated as 43.59 eV that corresponds to -43.59 V bias voltage to keep in secondary electrons inside the Faraday cup \cite{FC2}.
The absorber material of the Faraday cup was selected as Copper. The minimum necessary cup thickness can be predicted by calculating the proton and electron ranges inside the absorber.
The range of the proton beam in the copper was calculated by using SRIM program \cite {SRIM} and the range of the electron beam in copper was taken from the NIST ESTAR database \cite{ESTAR}. 20 keV proton beam ranges 0.1171 $\mu$m inside  copper and 10 keV electron beam which is the lowest energy of NIST ESTAR database that is much more higher than the our calculated secondary electron energy is found as 0.514 $\mu$m. However, a much larger Copper thickness was chosen,  8~mm, to make the assembly easier, and to distribute the heat load due to incident beam through the FC.
The secondary electron emissions were simulated by using IBSIMU and CST [\cite{ibsimu},\cite{cst}].

IBSIMU performs simulations independent from material and CST has different secondary electron emission models depending on the absorber material.  Secondary electron yield data for copper was taken from literature for a proton beam of 5 to 20 keV energy and imported in CST for simulation studies\cite{FC7}. 

It was planned to use the same Faraday cup in different locations of the proton line which are at the exit of the ion source, in the Measurement box and at the end of the LEBT line. Therefore to account for the the possibility that the beam diameter may be different at different locations of proton line, 
simulations with pencil beams with diameters varying from 10 to 40 mm with a 5 mm increments were performed.
 Backscattered protons were not taken into account in these simulations.
 
\begin{figure}[!htbp]
\centering
\includegraphics[scale=0.7]{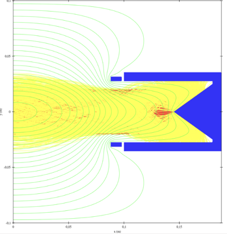} \includegraphics[scale=0.7]{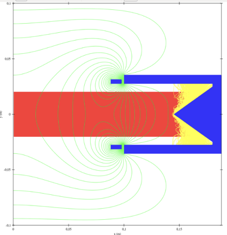}\\
\includegraphics[scale=0.7]{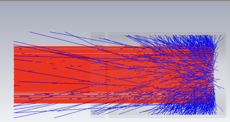} \includegraphics[scale=0.7]{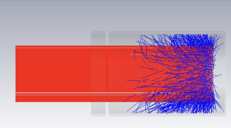}

\caption{FC simulations with different software and bias voltages. Top figures are from IBSIMU and the bottom ones from CST. Left simulations have no bias voltage applied and the right ones have -200~V. }
\label{fig:FCsimus}
\end{figure}

Proton backscattering from the copper surface was examined by using TRIM code \cite {SRIM} and it was found out that backscattered protons are increasing with increasing beam angle with respect to surface. A conic shape of 35 degree cone angle was chosen to limit the current reading errors related with backscattered protons.  
This setup was found to correspond to 5.23\% backscattered protons for a 20 keV incident pencil proton beam.  
The maximum percentage of the backscattered protons should be limited as it constitutes a source of error.
To retain most of these backscattered protons in the Faraday cup, various scenarios have been considered 
in IBSIMU and CST simulations, and the design has been optimized accordingly.
Simulation results of IBSIMU and CST can be seen for different biasing voltages in Figure \ref{fig:FCsimus}. 
Results from these two independent software programs are in agreement with each other.

\begin{figure}[!htbp]
\centering
\includegraphics[scale=0.27]{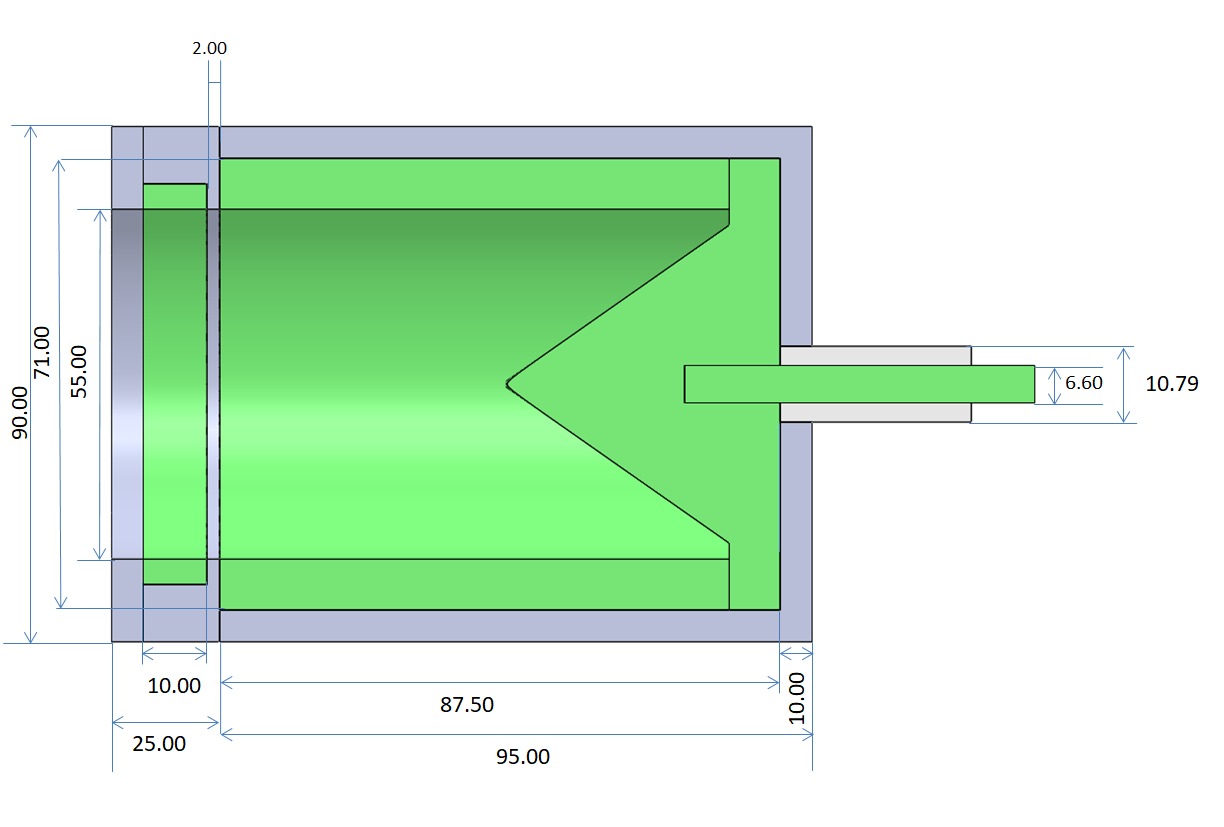}
\caption{Design of the Faraday Cup, all units are in mm}
\label{fig:FCdesign}
\end{figure}

The final  design of the FC can be seen in  Figure \ref{fig:FCdesign}.
As the average beam power is expected to be in the order of 4~W, a cooling system was not considered for this FC. 
Additionally its multiple planned locations in the beamline
would make multiple cooling setups. Therefore, this FC is designed for making measurements during brief periods.  
The FC is constructed using copper of 8~mm thickness, with an inner radius of 27.5 mm and length of 120 mm including the bias ring. 
Faraday cup and the bias ring are separated electrically
using a Teflon disc of 2~mm thickness. 
In the final setup the FC is to be at ground and the bias ring voltage can vary between -50~V and -200~V. 
Also a Faraday cup vacuum cover was designed to use Faraday cup outside the Measurement box. FC's copper body was covered with Teflon to prevent the contact with metal surfaces of the MBOX and also with the vacuum cover. 
During data taking the FC output, at the right side of Figure \ref{fig:FCdesign}, was connected to an oscilloscope over a 10~kOhm resistor for beam current measurements.

\subsubsection{Scintillation Screen}
The scintillation or phosphorus screens (SS) provide a direct method for observing and recording transverse beam profiles along the accelerator beam lines. 
The single image captured on these screens provide both X and Y profile information and when coupled with the Pepper Pot plate it yields beam emittance.
The two most commonly used phosphor screen types for image intensifiers are P43 and P46.  Surface-coated phosphor screens are reusable and are not as sensitive as plastic scintillator screens: they can be repeatedly exposed to luminescence. The phosphorus screens are also preferred in terms of cost.

The bill of materials required for constructing a phosphorus screen is relatively simple: a glass base, fluorescent powder, and isopropyl alcohol (99.9\% purity). The fluorescent powder can be obtained from pre-made fluorescent lamps such as Philips Master TL-D840 and TL-D830. These lamps are preferred due to the type of phosphor coating used on their surfaces. The warm white and cool white phosphor colors have a color temperature range of 3000-4000 K. The spectral irradiance values align with a standard Bell curve, centered around 530-630~nm. These wavelength ranges correspond to those needed in the beam line.

\begin{figure}[!htbp]
\centering
\includegraphics[scale=0.14]{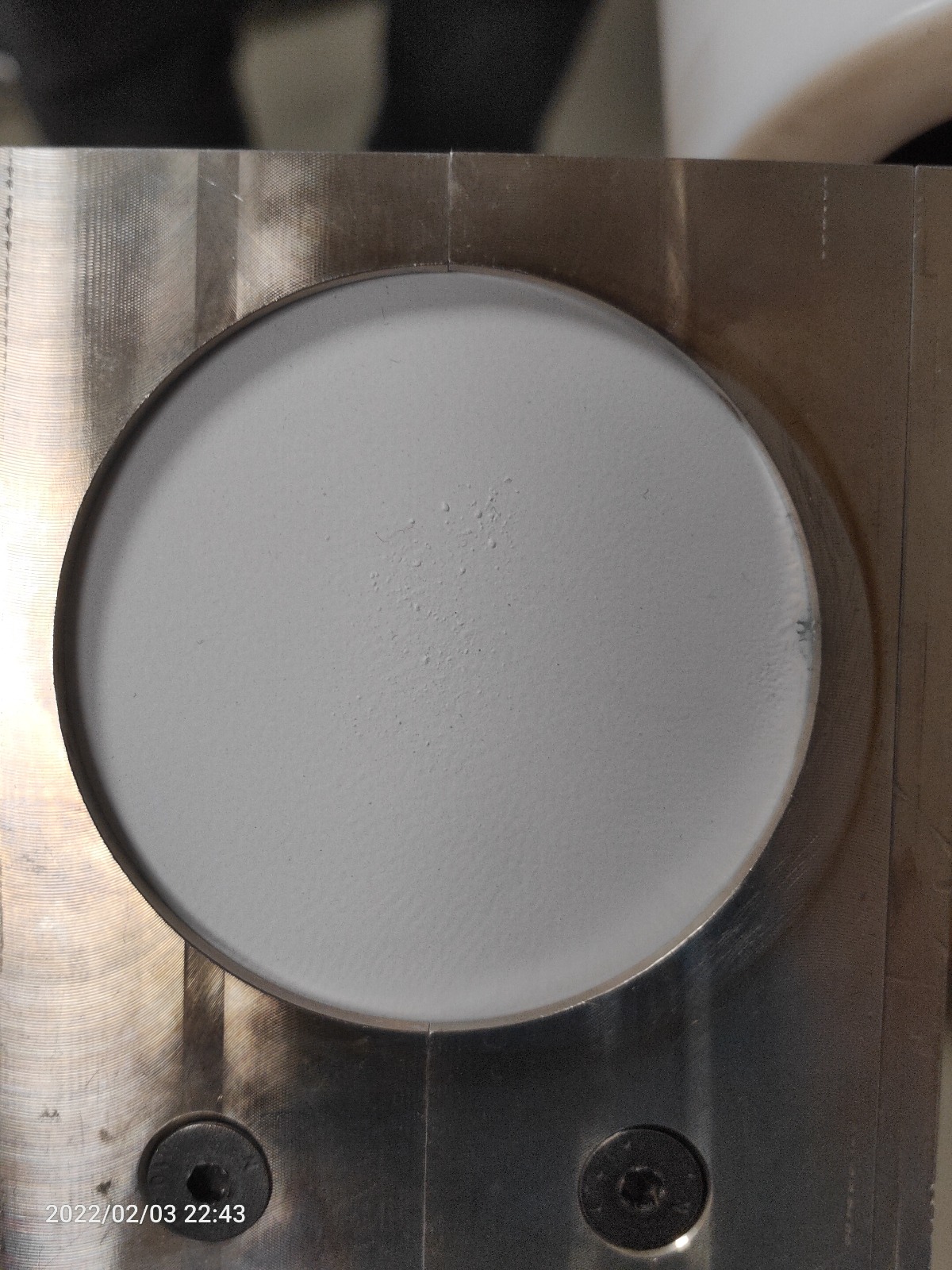} $\quad$
\includegraphics[scale=0.73]{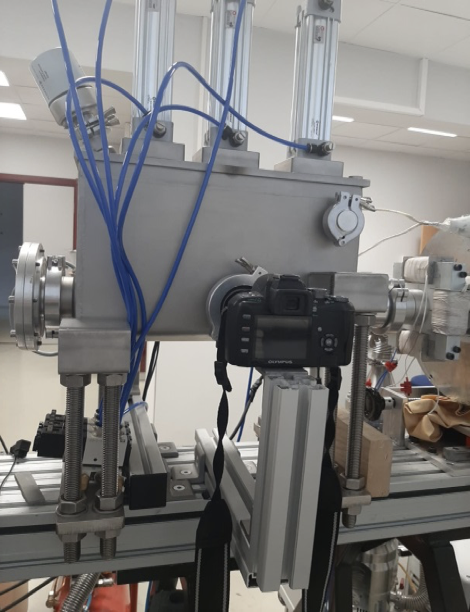}
\caption{Left: A home made Fluorscent Screen, Right: Beam diagnostics data taking setup}
\label{fig:camera}
\end{figure}

The final scintillation screen was therefore made locally in the laboratory, following a well defined procedure, on a 300~$\mu$m thick glass square of $60\times60$ mm using fluorescent powder. Such a detector can be seen in Figure \ref{fig:camera} left side before its installation in the MBOX.
A mirror, mounted at a 45-degree angle behind the phosphorus screen, projects the image of the beam through the vacuum window into the camera. The data taking setup can be seen in Figure \ref{fig:camera} right side. During the experiments the external lights are turned off to eliminate background noise light to the camera. 
This detector setup is also used in the emittance measurement which will be discussed in the next subsection.

\subsubsection{Pepperpot Plate}
\begin{figure}[!htbp]
\centering
\includegraphics[scale=0.23]{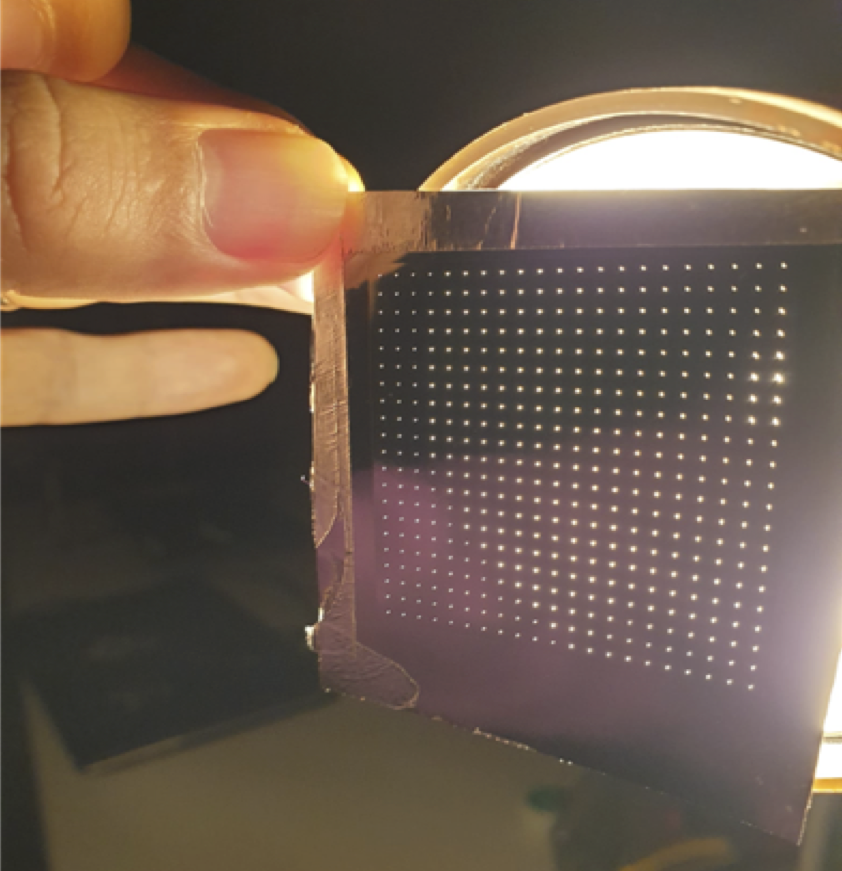}
\includegraphics[scale=0.80]{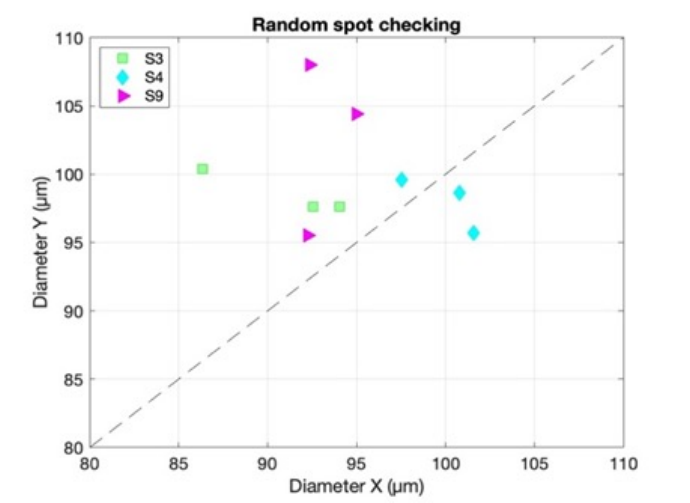}
\caption{Left: a PP sample with 21x21 hole array, Right: Random diameter check for 3 different samples}
\label{fig:RandomSpots}
\end{figure}

The main components of the emittance measurement are the pinhole (pepperpot) plate, a phosphorus screen, a plane mirror at 45 degree angle, and a camera attached to the vacuum window. 
The location of the PP with respect to the SS, the number, spacing and the diameter of the holes were optimized using python simulations discussed in the measurements and analysis section. 
The optimization goal in the simulated experiments was to reduce the emittance measurement error. 
Although the ideal pinhole plate material is tungsten,
a $250~\mu m$ thick stainless steel was selected for its lower cost of mask production.
21x21 Pinholes with a diameter of about 100 $\mu$m are spaced 2~mm horizontally and vertically and cover an area of $50 \times 50$~mm. 
The holes are made by the use of a fiber laser producing bursts of pulses at 50 GHz repetition rate (20 ps in between subsequent pulses). The burst duration and pulse energy were varied but roughly 10~nJ and 100~ns . The burst repetition rate was about 1~MHz. The pulse duration was in the range of 100-200~fs. 
\footnote{ The authors would like to thank the UFO lab, Bilkent University, Ankara, Turkey for their invaluable contribution in building the pepperpot mask.}
The uniformity of the hole diameter has been checked on various production runs to optimize the laser parameters. Figure \ref{fig:RandomSpots} contains such control measurements for 3 different samples. For the final production, the so called "sample-4" has been used as it has the most uniform diameter distribution. 
In order to prevent thermal deformation of the mask, it is sandwiched between two aluminum frames of 500 $\mu$m thickness each. These support frames are used to prevent any deformations that may occur on the surface of the perforated plate. The distance between the SS and the the pepper pot mask is set as 12.35 cm after simulations meanwhile that distance can be arranged by simulations for less or more divergent beams since the mask is mounted on a rail in order to be slide.

\subsection{Control and Readout}

\begin{figure}[!htbp]
\centering
\includegraphics[scale=0.30]{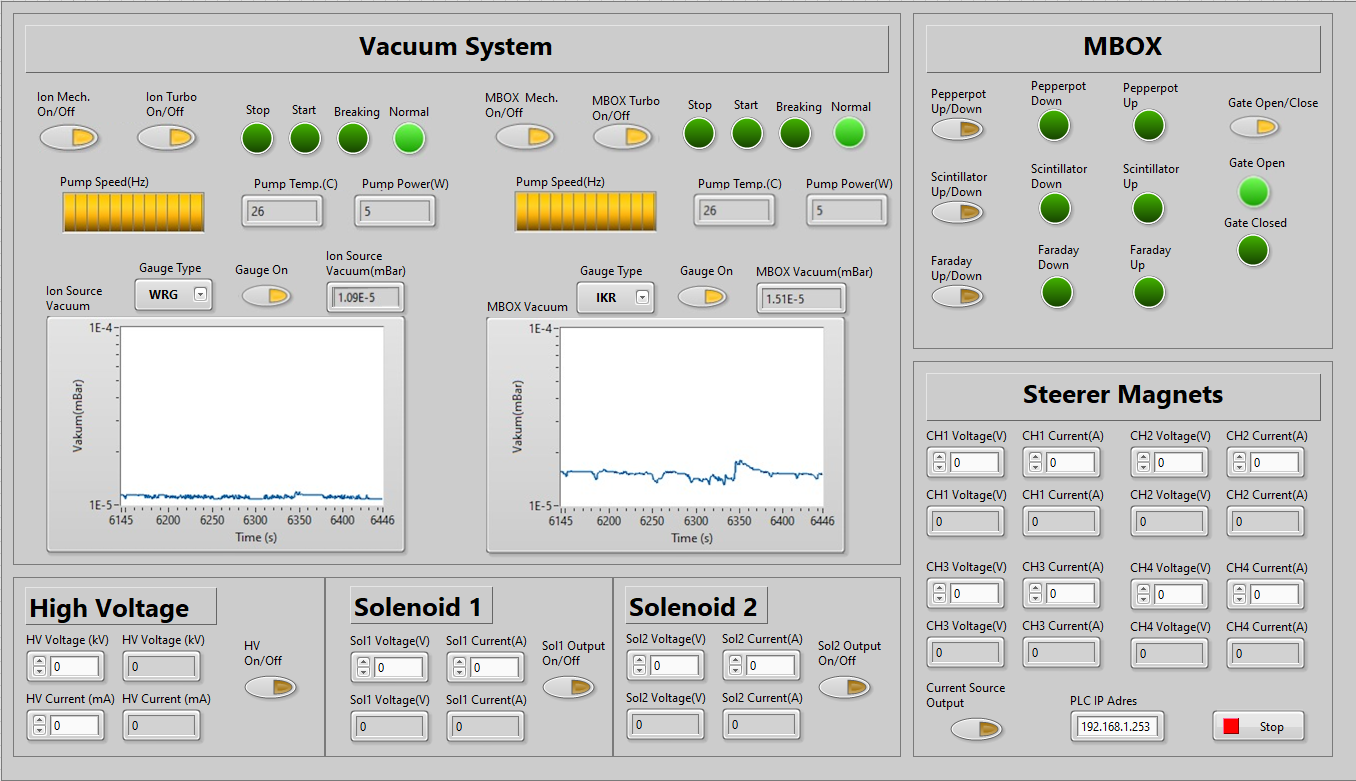}
\caption{Control and Monitoring GUI for the LEBT line and MBOX}
\label{fig:Controls}
\end{figure}

The KAHVELAB proton beamline control and monitoring system is responsible a high-voltage power supply, four current sources, two mechanical vacuum pumps, two turbo-molecular vacuum pumps, various vacuum gauges, the vacuum gate valve separating the MBOX from the upstream of the beamline, temperature sensors and three pneumatic actuators devices.
In the first version of the setup, a PC running a LabVIEW program written in the G language was used to
control and monitor the equipment, while the low level hardware access was achieved using Arduino micro-controller cards \cite{ISLEBT-PTAK}.
Although the system worked in principle, there were stability issues with this  version: The LabVIEW  graphical user interface (GUI) was to sluggish to handle user interactions, the Arduino boards often lost their firmware allowing connection to LabVIEW etc.
To alleviate these problems a second version of the control and readout software is being written. 
In this new version of the setup, all the hardware devices (such as the vacuum pumps, vacuum gauges, the MBOX actuators etc) are now controlled and monitored by a Siemens S7-1200 1215C model PLC. 
The SP7 LabVIEW Toolkit auxiliary plugin is used to enable communication between the PLC and the PC.
The PC runs a new LabVIEW GUI which interacts with the user and  exchanges with the PLC. A screen shot from this GUI is given in the Figure \ref{fig:Controls}. 
As one can notice, the GUI displays pump status,  graphically represents pressure inside the MBOX and the ionization chamber, displays all temperature, current and voltage values and naturally the MBOX actuator positions.

\section{Beam Measurements and Analysis}
After being produced and subjected to vacuum training, the MBOX was installed at the KAHVELab proton line in early 2022, and subsequent beam measurements were conducted. 
In the next paragraphs  these measurements are presented and discussed.

\subsection{Beam Current}
\begin{figure}[!htbp]
\centering
\includegraphics[scale=0.50]{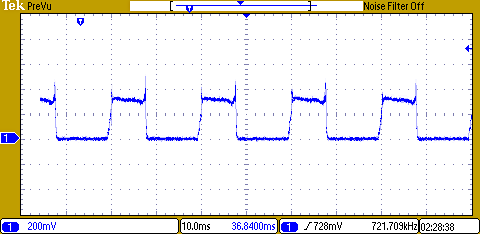}
\caption{Beam charge measurement using FC}
\label{fig:Charge}
\end{figure}

The beam current measurement is a destructive one achieved using the FC, lowered into the beamline using the relevant actuator.
An example from such a  measurement is presented in Figure \ref{fig:Charge}.
The signals are obtained using an oscilloscope that reads the FC located at z = 87 cm (see Fig. \ref{fig:LEBTLineMbox}), through a voltage divider circuit.  The measurement reveals a pulse width of 8 ms and a pulse period of 20 ms. Consequently, the duty factor can be calculated as the ratio of active time to the period which is 8/20 = 0.4. Similarly, the instantaneous current is determined to be 0.03 mA, since the relevant resistor in the voltage divider circuit is 10 kOhm. The average current is calculated at 0.012 mA, below the value reported by the high voltage power supply. 
This beam, with a lower current with respect to the design value of 1~mA, is obtained by tuning three parameters at the same time:
1) reduction of the magnetron input voltage using a manually controlled variac, 2) adjustment of the very first tuner after the magnetron 3)  reduction of the Hydrogen gas flow rate to a minimum of 0.01 sccm. 
The rational behind this ``pilot'' beam is to test the whole system using a small charge and, once verified, to increase the beam current gradually. 
The measurements with this increase are planned for after the completion of the RFQ installation. However since PIXE applications require a beam current of few nA, even this pilot beam would be sufficient to achieve this goal.

\subsection{Beam profile}
\begin{figure}[!htbp]
\centering
\includegraphics[scale=0.068]{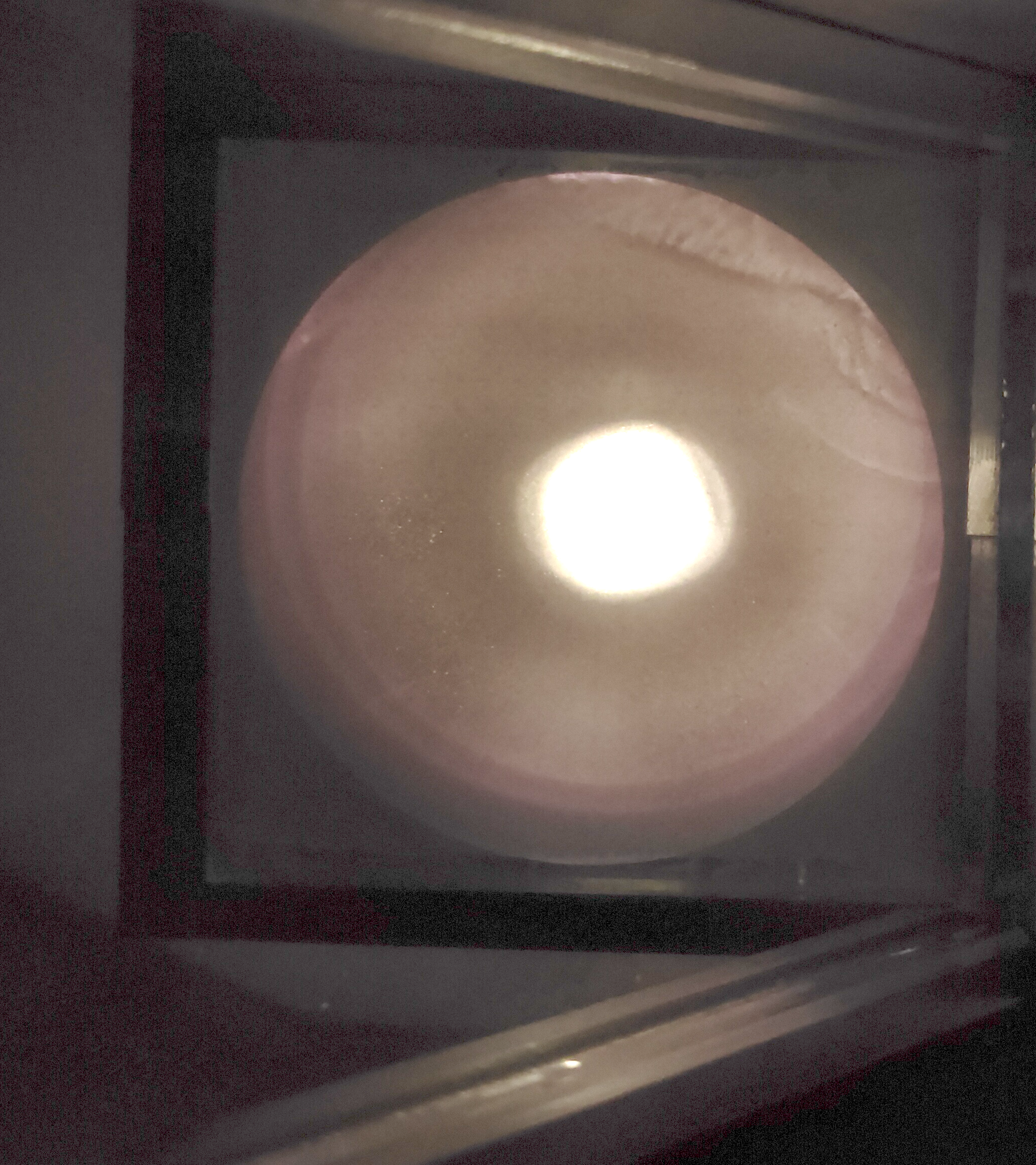}
$\qquad$
\includegraphics[scale=0.27]{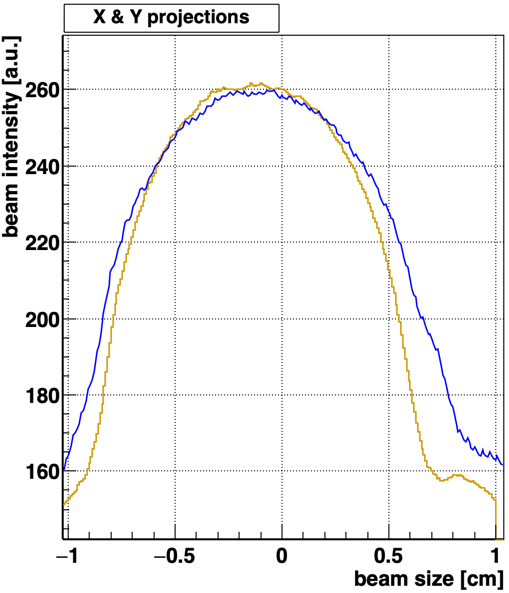}
\caption{Left: Raw proton beam profile photo at MBOX, Left: Beam profiles in X(blue) and in Y(orange)}
\label{fig:beamAtMBOX}
\end{figure}

For this measurement the SS and the attached 45 degree slanted mirror are lowered to intercept the beam. 
The image is recorded with a digital  camera
and is  processed later using a locally developed Python program. 
The raw recording format was selected to avoid any data loss which might arise from JPEG data compression algorithms. 
The Figure \ref{fig:beamAtMBOX} contains such an image on the left side.
This image is read in using the Python OPENcv library and converted
into a two dimensional brightness histogram from which X and Y projections are obtained. The calibration i.e. the conversion from pixel counts to length in mm is achieved by using the known lengths of the plane mirror frames. 
In  Figure \ref{fig:beamAtMBOX}, the right plot shows the beam profiles in the X (blue) and Y (orange) directions. 
It can be observed that the distributions are slightly off axis and the X has a larger beam width. 
As in the literature, the Full Width at Half Maximum (FWHM) is used to estimate the beam width: 
 14 and 13~mm are obtained for X and Y directions respectively. 
 The beam size, $\sigma$ is usually obtained from a fit to the beam profile, however the fitted
 Gaussian function undershoots and the hyperbolic secant function overshoots the data in various locations. 
 We therefore use the linear relation between the FWHM and $\sigma$ for these distributions by taking the average of the correlation constants, 2.355 for the Gaussian and 2.634 for the hyperbolic secant functions.
 The beam size estimation is therefore 5.6 (5.2) mm for X (Y) axes.
 A simulation of the proton beam in DemirciPRO, yields the expected RMS beam envelope at the SS as 5.1~mm in both directions. Given that DemirciPRO does not consider space charge effects, thus underestimates the beam size, one can conclude that the experimental results and simulations are consistent with each other.

\begin{figure}[!htbp]
\centering
\includegraphics[scale=0.18]{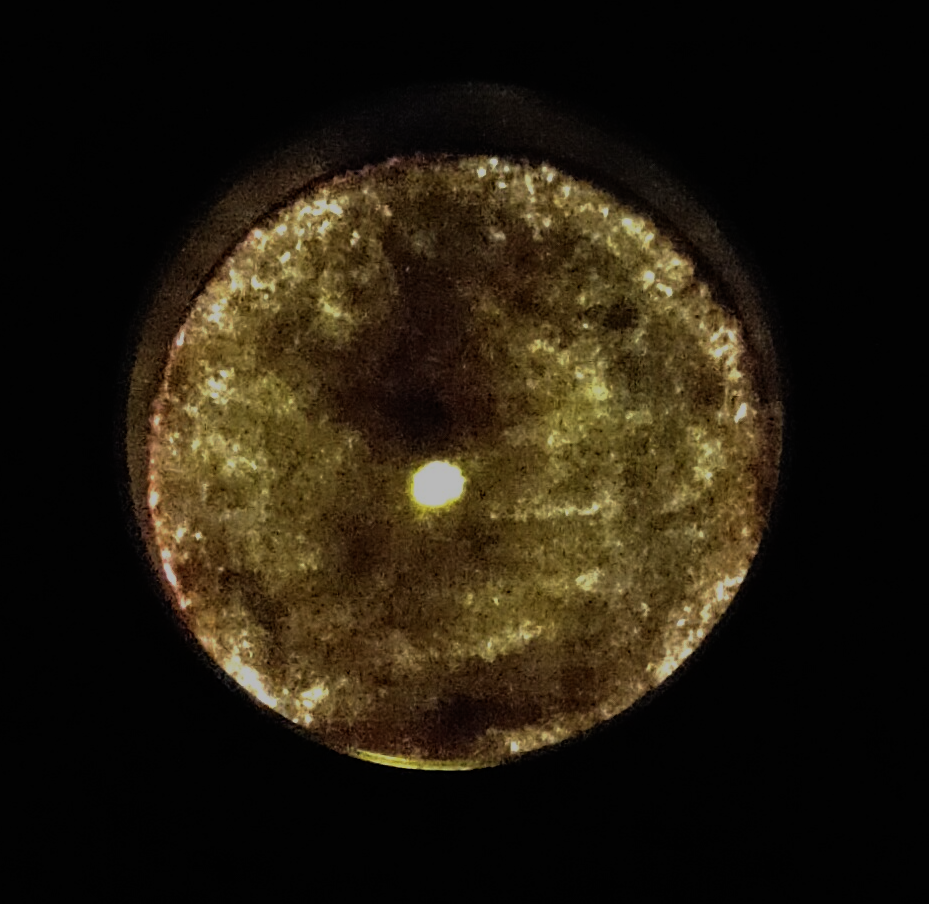}
$\quad$
\includegraphics[scale=0.138]{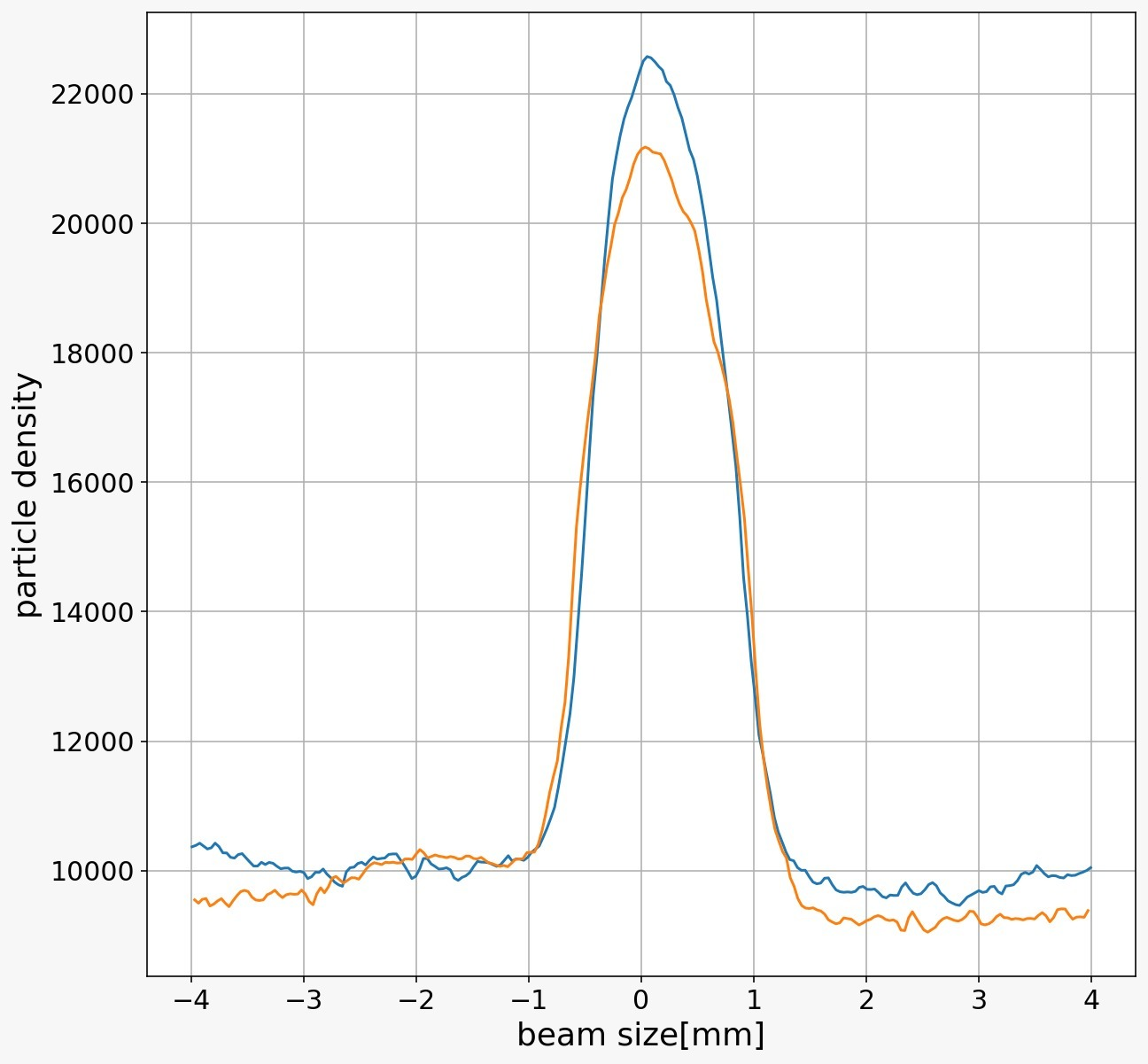}
\caption{Left: Raw beam photo taken at the end of the LEBT line, Right: X(blue) and Y(orange)  projections}
\label{fig:beamAtEnd}
\end{figure}

A similar setup was also used to obtain the beam profile of the focused beam at the end of the LEBT line. 
In this location the photo is taken without a mirror and is presented in Figure \ref{fig:beamAtEnd} left side together with the projection histograms on the right side. Notice that the maximum height of the signal is at the center position (0mm) as compared to the background noise originating from impurities on the Fluorescent Screen. 
However the beam is asymmetrical in both directions towards the positive side (up and right). The reason for this asymmetry is under investigation.   
The FWHM of the beam is measured as about 1.5 mm for both X and Y directions. The beam is focused by Sol-2 at the measurement location which represents the RFQ input.

\subsection{Beam emittance}
\begin{figure}[!htbp]
\centering
\includegraphics[scale=0.16]{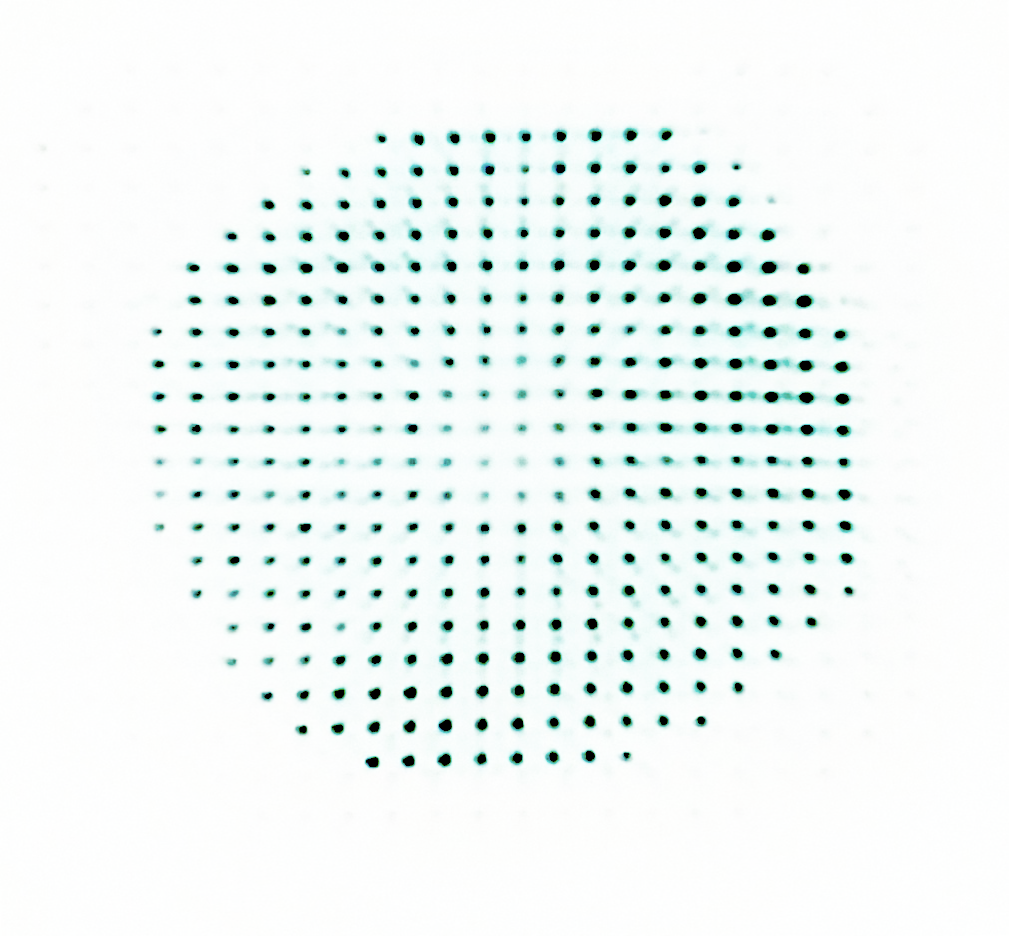}
\includegraphics[scale=0.13]{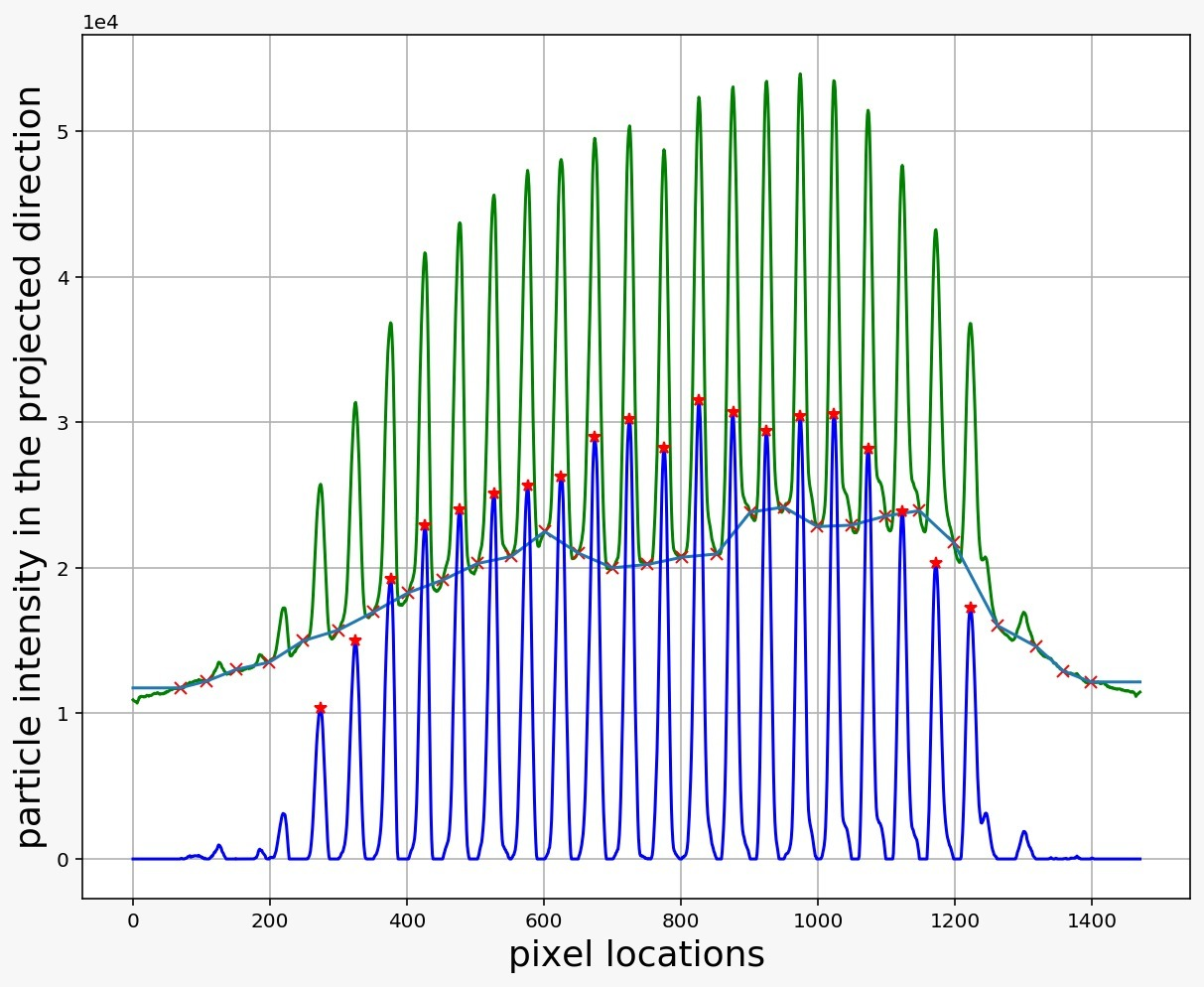}
\includegraphics[scale=0.11]{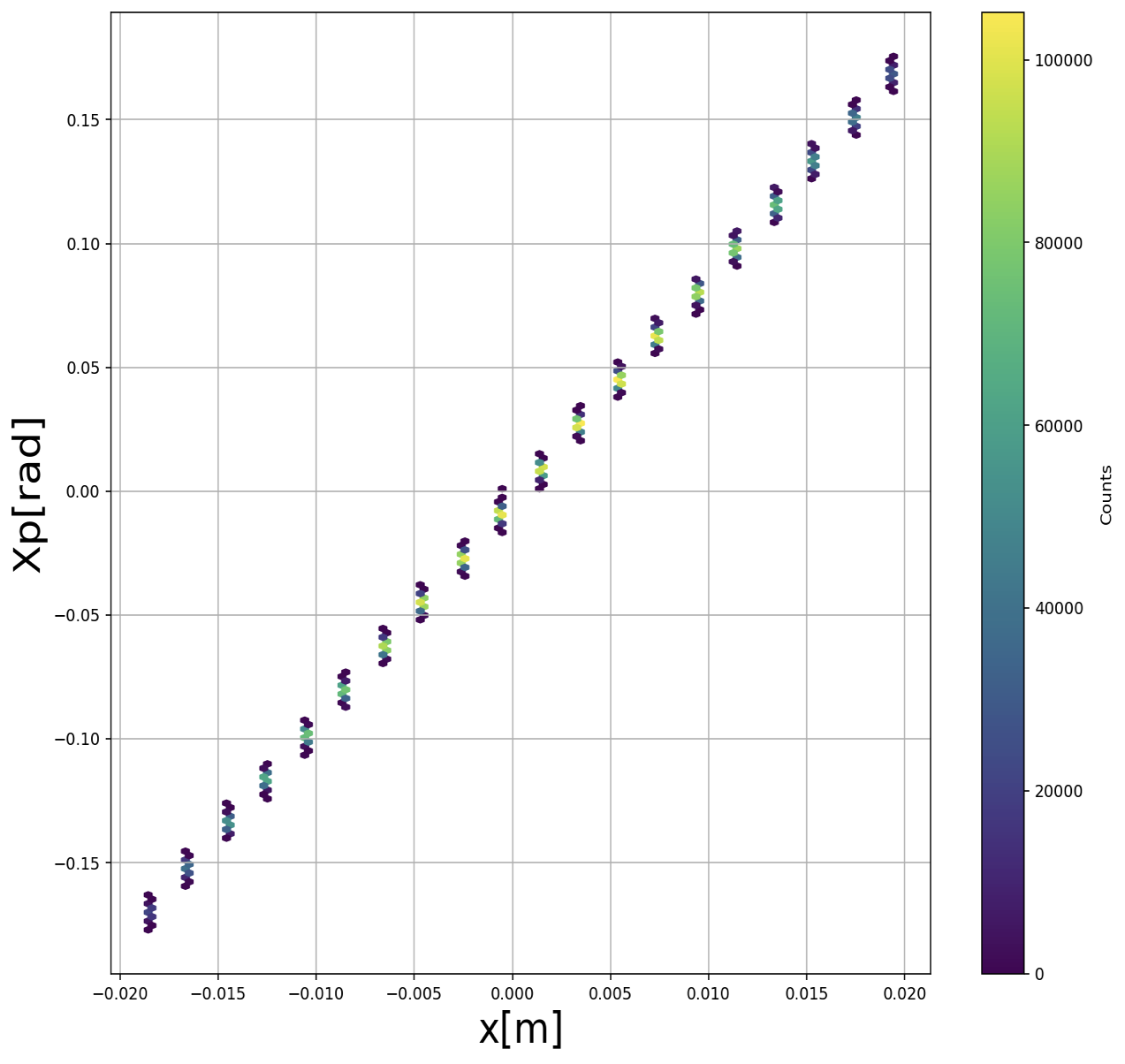}
\caption{Left: Intensity on the SS after PP, Center: The beamlet distribution before (green) and after (blue) background subtraction, Right: the emittance graph in X direction}
\label{fig:PPPanalysis}
\end{figure}

The beam emittance measurement is performed by lowering both the PP plate and the SS into the beamline. As previously discussed, the mask allows only a small portion of the proton beam to reach the SS.
The image formed by the scintillation light of the particles surviving the PP is recorded by the camera. Such a photograph, after color inversion, can be seen in Figure \ref{fig:PPPanalysis} left side. 
The image file is read in using Python OPENcv library and converted into a two dimensional intensity histogram.
The X and Y projections of the intensity histogram yield the 
beamlet intensity distributions, also containing some background light.  
There are multiple methods to estimate and subtract the background and to find the peak positions of the beamlets. 
The background finding method applied in this note is to calculate the dip position equidistant to two adjacent peaks   and define a segment of line between consecutive peaks. 
These line segments define the background that will be subtracted to find the signal distributions. 
In Figure \ref{fig:PPPanalysis} center plot shows such an example.  
The original distribution is shown in green, the estimated background is shown with the purple line and resulting signal only distribution is drawn in blue. 
Although there are multiple true beamlet candidates, only 21 of those should be considered as there are 21 holes per dimension in the PP. The procedure is to consider the central and brightest peaks, marked with a red star in the same image.
Therefore the center position of the hole through which protons pass, is accepted as the position of the relevant beamlet. The simulated macro particles with coordinates $(x_i, y_i)$ will pass through the pepperpot mask where each pinhole located at $(x_j^{ph}, y_k^{ph})$, if the individual macroparticle yields:
\begin{equation}
    (x_i-x^{ph}_j)^2 + (y_i-y^{ph}_k)^2 < r^2.
\end{equation}
Then the assumption for the simulated macroparticle having the coordinate $(x_i, y_i)$ would be having the coordinate of the pinhole center $(x_j^{ph}, y_k^{ph})$, then the angle associated to that macroparticle would be calculated as:
\begin{equation}
   x'_{i, new} = arctan(\frac{x_i + x'_i L}{L}), 
\end{equation}
which is valid for the vertical counterpart, as well. 
The marked beamlets therefore yield the divergence angles since the distance between the PP and SS is known, i.e. L=123.5~mm.
Using the angles ($x'_{i, new}$) and positions ($x_j^{ph}$),  the geometrical emittance and the Twiss parameters can be substituted into the rms emittance formula:
\begin{equation}
   \epsilon_{x, g}^2 = <(x^{ph}_j)^2> <(x'_{i, new})^2 > - < (x_j^{ph}) (x'_{i, new})>^2.
\end{equation}

The result from a calculation in the X direction is shown in Figure \ref{fig:PPPanalysis} right side.
Using the same beamline settings as in the previous section, the normalized emittance in the X (Y) direction is found to be 0.029 (0.033) $\pi$mm.mrad .
This value is compatible with the IBSIMU expectation of 0.0297 $\pi$mm.mrad in both directions. 
The same analysis also yielded the Twiss parameter $\beta$ as 6.63 (5.75) $mm/{\pi}mrad$ in the X (Y) direction. 
These values are in agreement with the predicted value of  6.5  $mm/{\pi}mrad$ obtained in a simulation with 1M particles created by  IBSIMU and tracked in the LEBT line by DemirciPRO which does not incorporate space charge effects. 
Another cross-check can be performed using the beamsize ($\sigma$) estimated in the previous section and the measured geometrical emittance ($\epsilon_g$).
The geometrical emittance value can be easily found since the proton beam energy is known as 20~keV.
Using the relation $\sigma^2 / \epsilon_g = \beta$, the Twiss parameter beta can be estimated as $7.05$ for X and $5.35$ $mm/{\pi}mrad$ for Y directions. These estimations are compatible with the direct measurements using the PP with a relative error of about 10\% level. 
Moreover, their average aligns with the simulation expectation value with an error better than 5\%. Given the fact that only a small portion of the beam survives the PP mask, the compatibility level across different measurements and simulations is remarkable.

The image analysis, emittance and Twiss parameter calculation software was initially developed locally using a beam file, created with IBSIMU program. The program is written in Python and does also beam tracking along the LEBT line, without considering the space charge effects. The LEBT parameters are given in a simple text based  configuration file. 
In this software, the position-angle vectors of the beam particles were transported along the LEBT line using the transfer matrices of the drift and solenoid spaces.
The PP is also introduced into this program and the particles surviving it were identified and later used to calculate the beam emittance on the SS. The results from this code were compared to the
output from the DemirciPRO program and found to be mutually compatible. 
One of its major advantages over DemirciPRO is the ability to directly work with image files. Further details of this software can be found elsewhere \cite{duyguMS}.

Multiple photos were taken at different current values of Sol-1 and analyzed as described above. These results, together with expectations from IBSIMU and DemirciPRO simulations are presented in Table \ref{tab:results}.
As shown, the solenoid current is scanned between 17.3 and 18.3 Amps corresponding to a 0.01~T field change. 
For lower and higher currents, the pepper pot method looses its applicability:
For higher currents, the beam becomes over focused, whereas for lower currents the beam becomes too divergent and beam losses occur at the beampipe walls. 
The average values of the normalized emittance in
both X and Y directions,
along with the corresponding standard deviations, are also presented in the same table. 
As one can notice the average emittance value agrees with the expectation within 1 (2) sigma in X (Y) direction. 
These measurements can be further improved by a better estimation of the light background and a faster camera.
Nevertheless, despite the home-made nature of the detectors
results of significant merit are obtained.

\begin{table}[!hbt]
\centering
\caption{ Emittance values simulated and calculated at different Sol-1 current values.}
\begin{tabular}{c|c|c}
	\textbf{I$_{Sol-1}$ (A)} & \textbf{$\epsilon$ X ($\pi$mm.mrad)}  
   &\textbf{$\epsilon$ Y ($\pi$mm.mrad)} \\ \hline
    17.3 & 0.039	& 0.061\\
    17.5 & 0.038	& 0.041\\
    18.0 & 0.036	& 0.041\\
    18.2 & 0.030	& 0.035\\
    18.3 & 0.029	& 0.033\\ \hline
   \textbf{Average}  & 0.0344	& 0.0392\\ 
   \textbf{Std.Dev.}  & 0.0046	& 0.0052\\ \hline
\textbf{Expected} & 0.0297	  & 0.0297\\ 
\end{tabular}
\label{tab:results}
\end{table}

\section{Conclusions}

In this report, we have introduced a compact and cost-effective beam diagnostics station designed for ion beams. The station includes detectors for beam current, profile, and emittance measurements, all of which have been locally designed and constructed. Additionally, we have developed our own control and data analysis software programs.
The entire setup has undergone successful testing using the low-energy proton beamline at the KAHVE Laboratory in Istanbul, Turkey. The values obtained for the proton beam current, profile and emittance were in agreement with simulations and with crude readings from other professional instruments. 
These reported measurements were done on the beamline without the RFQ and with the MDIS which used electromagnets.  
While the RFQ was being manufactured, the MDIS 
was upgraded to use permanent magnets. This upgrade removed the cooling and EM solenoid requirements and simplified the setup. The commissioning of the new ion source is ongoing. 
The production of the two RFQ modules is also completed and the RFQ is assembled.
Its vacuum and electromagnetic tests are ongoing. 
The first runs of the proton beam line including the RFQ is scheduled for Q4 of 2023.

We plan to use the MBOX, in the coming years, at the KAHVELab proton beamline with higher beam currents and with possible minor upgrades such as the further automation of the data taking.
Given our satisfaction with its performance, we particularly recommend this approach of developing complete, in-house solutions for beam diagnostics to educational laboratories, rather than purchasing a ready-made system with off-the-shelf components.
This approach not only supports local industry but also provides valuable hands-on experience for physics and engineering students. One significant advantage of this approach is its low cost. However, it is worth mentioning that building a measurement station from scratch does require time and effort, as opposed to acquiring readily available components. Nonetheless, the benefits of local manufacturer development and student involvement make it a worthwhile endeavor.

\section*{Acknowledgements}
The main project is being supported by by T\"{U}B\.{I}TAK grant 119M774, as well as \.{I}stanbul University BAP grant 33250 and 36823.
The authors would like to thank the UFO lab, Bilkent University, Ankara, Turkey for their invaluable contribution in building the pepperpot mask.

\end{document}